\documentclass[
    ,final            
  ]
  {aipproc}

\layoutstyle{6x9}

\begin{document}

\title{The magnetic fields of accreting T Tauri stars}


\author{S. G. Gregory}{
  address={SUPA, School of Physics and Astronomy, Univ. of St Andrews, St Andrews, KY16 9SS, UK},
  ,email={sg64@st-andrews.ac.uk},
}
\author{S. P. Matt}{
  address={Dept. of Astronomy, Univ. of Virginia, P.O. Box 400325, Charlottesville, VA 22904-4325, USA},
  altaddress={NASA Ames Research Center, Moffett Field, CA 94035-1000, USA},
}
\author{J.-F.  Donati}{
  address={LATT - CNRS/Universit\'{e} de Toulouse, 14 Av. E. Belin, F-31400 Toulouse, France}
}
\author{M.  Jardine}{
  address={SUPA, School of Physics and Astronomy, Univ. of St Andrews, St Andrews, KY16 9SS, UK}
}


\begin{abstract}
Models of magnetospheric accretion on to classical T Tauri stars often assume that the stellar magnetic field is
a simple dipole.  Recent Zeeman-Doppler imaging studies of V2129~Oph and BP~Tau have shown however that
their magnetic fields are more complex.  V2129~Oph is a high mass T Tauri star and despite its young age is believed
to have already developed a radiative core.   In contrast to this, the lower mass BP~Tau is likely to be completely
convective.  As the internal structure and therefore the magnetic field generation process is different in both stars, it is 
of particular interest to compare the structure of their magnetic fields obtained by field extrapolation from magnetic 
surface maps.   We compare both field structures to mulitpole magnetic fields, and calculate the 
disk truncation radius for both systems.  We find that by considering magnetic fields with a realistic degree of 
complexity, the disk is truncated at, or within, the radius obtained for dipole fields. 
\end{abstract}

\maketitle


\section{Introduction - Mapping stellar magnetic fields}
Classical T Tauri stars (cTTs) are young pre-main sequence stars which accrete material from circumstellar disks.  
Many observations support the magnetospheric accretion scenario, where the stellar magnetic field is 
strong enough, and sufficiently well-ordered, to truncate the disk at a few stellar radii \cite{kon91}.  Gas is forced to follow the
field lines of the stellar magnetosphere and accretes on to the star at high velocity, producing detectable hotspots \cite{edw94}.
However, until recently little was known about the geometry of their magnetic fields, with the majority of models
assuming a dipole.

Measuring the Zeeman broadening 
of unpolarized spectral lines has proved to be very successful in detecting T Tauri magnetic fields (e.g. \cite{joh07}),
although this technique is insensitive to the magnetic topology.  In contrast, measuring the 
polarization signature in spectral lines gives access to the field topology, and allows information about how the 
magnetic energy is distributed within the different field components to be determined.  However, like all 
polarization techniques, this suffers from flux cancellation effects and yields limited information
regarding the field strength.  

\begin{figure*}
        \centering
        \begin{tabular}{cc}
                 \label{v2129oph_map}      
                 \includegraphics[height=.22\textheight]{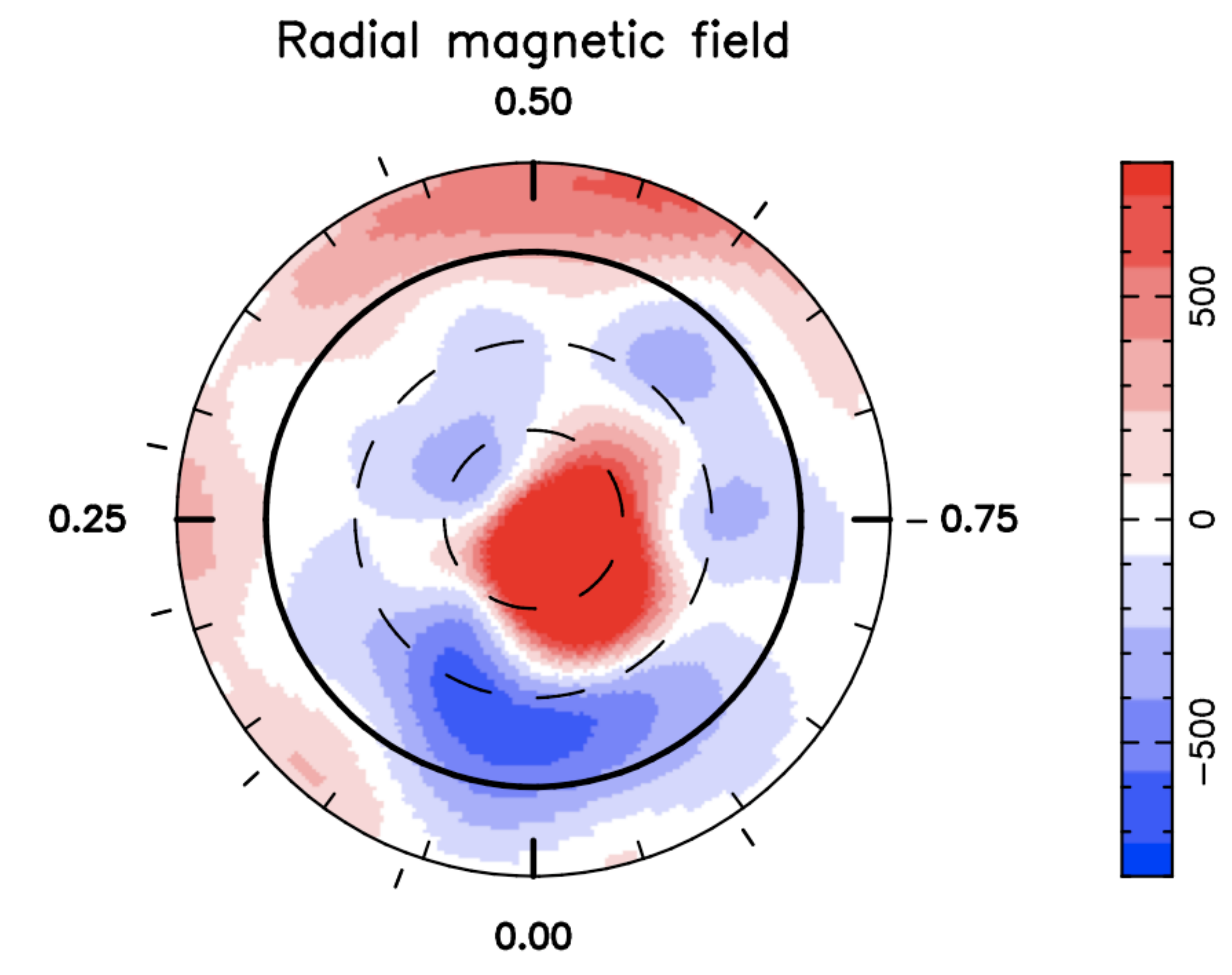}
                        &
                 \label{bptau_map}
                 \includegraphics[height=.22\textheight]{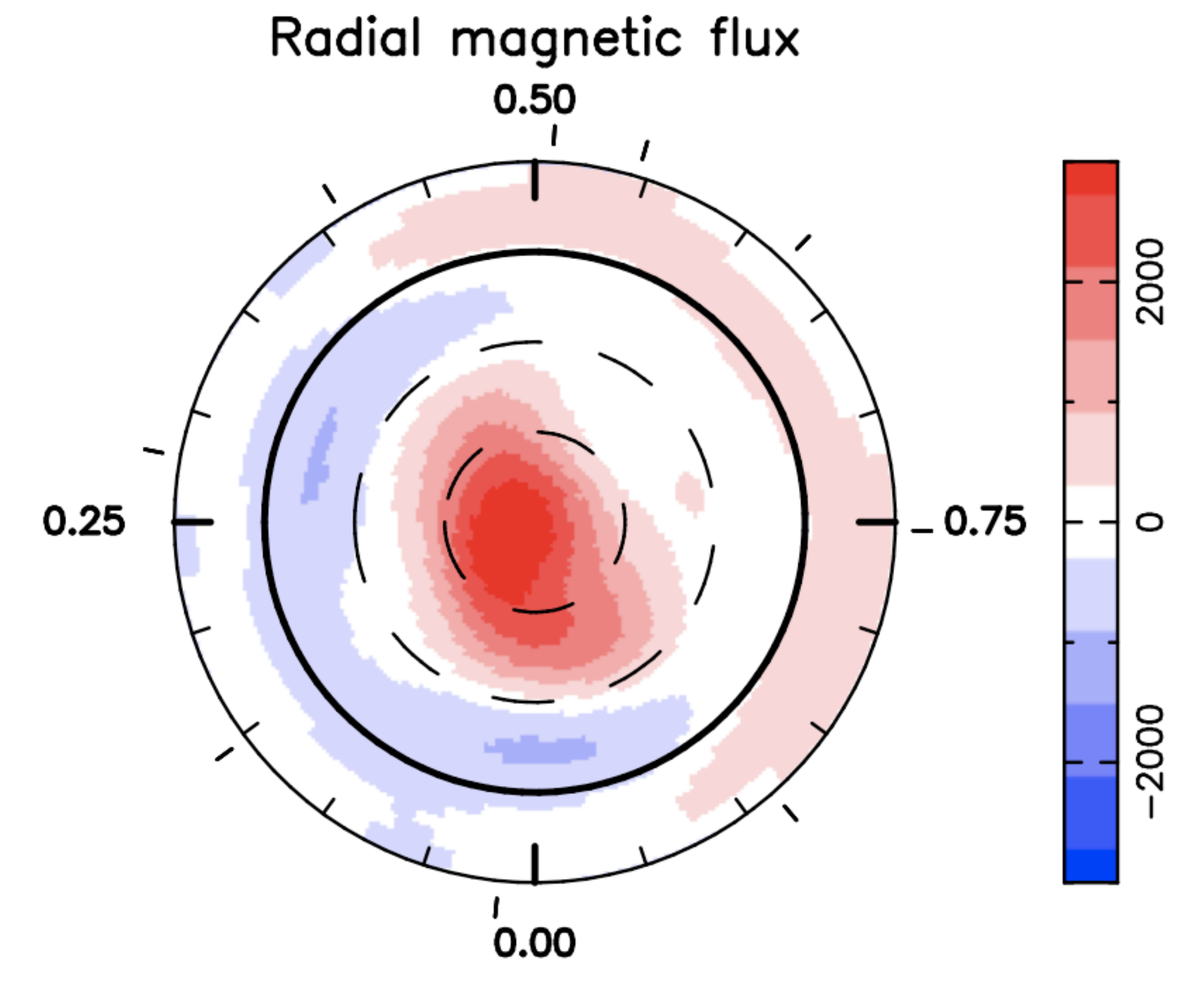}
                          \\
        \end{tabular}
        \caption{Flattened polar projections of V2129~Oph (left) and BP~Tau (right) showing the strength in Gauss and polarity of the radial
               field component \cite{don07,don08}.  Red denotes positive, and blue negative, field regions.  The solid black circle represents 
               the stellar equator, and the dashed circles lines of constant latitude.  The surface magnetic fields are clearly non-dipolar.}
        \label{maps}
\end{figure*}

The polarization signatures detected in spectral lines are typically
small, and therefore cross-correlation techniques (such as Least-Squares Deconvolution \cite{don97}) were
developed in order to extract information from as many spectral lines as possible.  The signal-to-noise ratio of 
the resulting average Zeeman signature is several tens of times larger than that of a single spectral line.   
Magnetic surface features produce distortions in
the Zeeman signature that depend on the latitude and longitude of the magnetic region, as well as the orientation of
the field lines.  By monitoring how such distortions move through the Zeeman signature as the star rotates, a method 
referred to as Zeeman-Doppler imaging (ZDI), the distribution of magnetic polarities across surface of cool stars can be 
determined (see Donati et al, these proceedings).  

ZDI studies have now been carried out on two cTTs, V2129~Oph and BP~Tau 
\cite{don07,don08}.  Both stars have complex non-dipolar magnetic fields (Fig. \ref{maps}).  The field of V2129~Oph is 
dominated by a 1.2~kG octupole field component, tilted at $\sim 20\,^{\circ}$ with respect to the 
stellar rotation axis.  The dipole component was found to be weak, with a polar strength of only 0.35~kG and tilted at 
$\sim 30\,^{\circ}$ \cite{don07}.  The field of BP~Tau consists of strong dipole and octupole field components of strength
1.2 and 1.6~kG, both tilted by $\sim 10\,^{\circ}$ with respect to the stellar rotation axis, but in different planes \cite{don08}.


\subsection{Field extrapolation and comparison with multipoles}
From the surface maps of the photospheric field, we reconstruct the
coronal field via field extrapolation, using the potential field source surface 
method \cite{alt69}.
This has the advantages over full MHD models of computational speed and simplicity, and does
not require assumptions about an equation of state, or the energetics.  The disadvantages are that
non-potential and time dependent effects can not be modeled.  However, it has been demonstrated
that the potential field model is often adequate and produces results which match those of
more complex MHD models \cite{ril06}.  We note that for cTTs the main source 
of non-potentiality in the field is likely to be due to the interaction between the stellar magnetosphere 
and the disk.  We do not consider this here - the field structures that we present are static 
and do not evolve in time.  

\begin{figure}
  \includegraphics[height=.29\textheight]{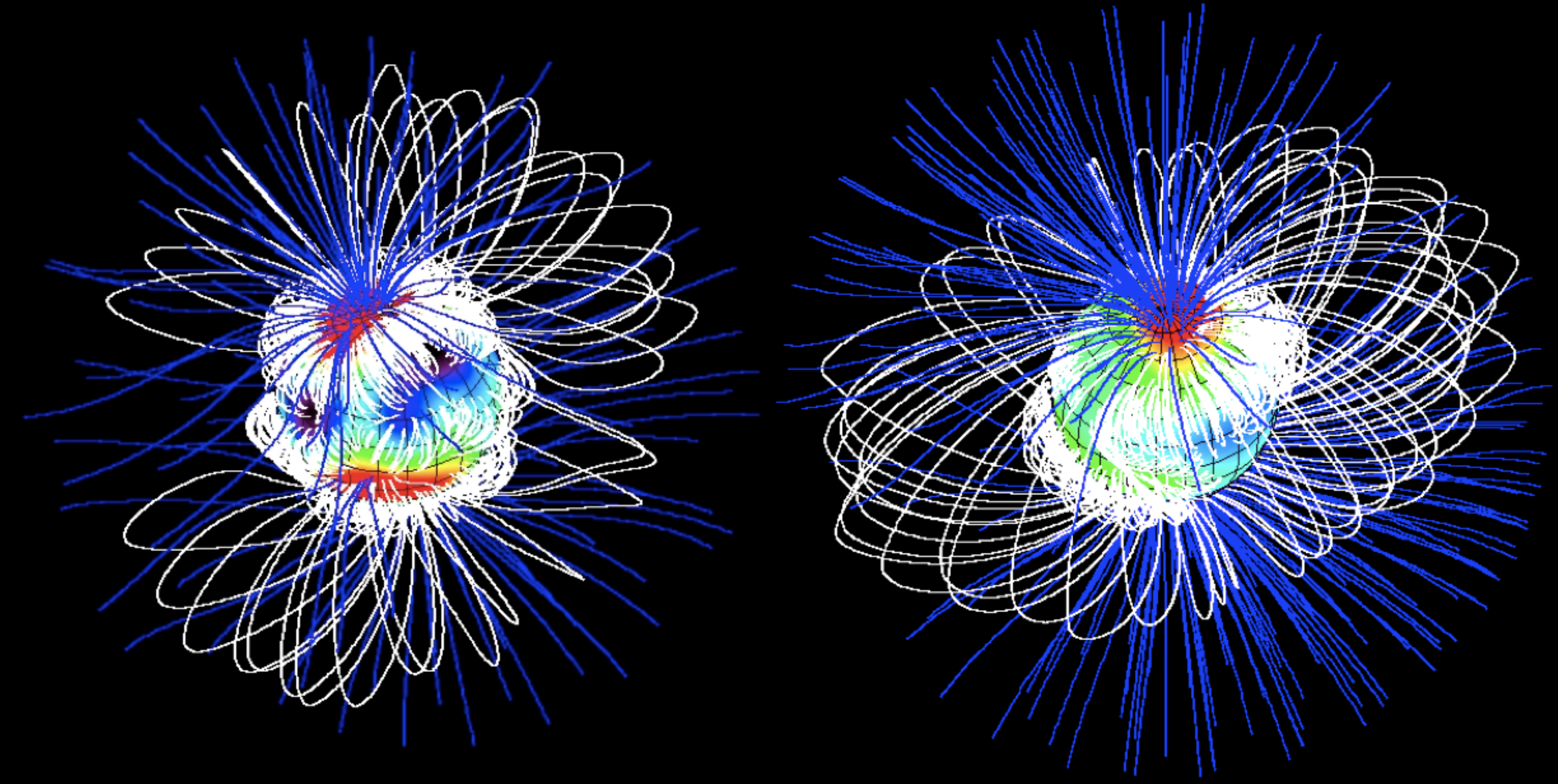}
  \caption{The field structure of V2129~Oph (left) and BP~Tau (right) derived by field extrapolation from the magnetic 
           surface maps in Fig. \ref{maps}.  Closed/open field lines are shown in white/blue.  There is a clear 
               distinction between the complex surface field and the simpler larger scale field.}
  \label{extrap}
\end{figure}

The potential field source surface model requires two boundary conditions.  The first is that the radial 
component of the magnetic field at the stellar surface is that which is determined from the Zeeman-Doppler 
maps.  The second is that at some height above the star, known as the source surface $R_S$, the field becomes
purely radial, which mimics the effect of a stellar wind blowing open field lines at some height above the star.  
Although the choice of $R_S$ is a free parameter of the field extrapolation model, its location is well constrained by the 
observed accretion hotspot locations \cite{don07,don08}.  Once $R_S$ has been set, predictions can then be made regarding the X-ray emission
properties of the coronal field \cite{jar08}, which will be tested against future X-ray observations.

\begin{figure*}
        \centering
        \begin{tabular}{cc}
                 \label{v2129oph_height_width}      
                 \includegraphics[height=.28\textheight]{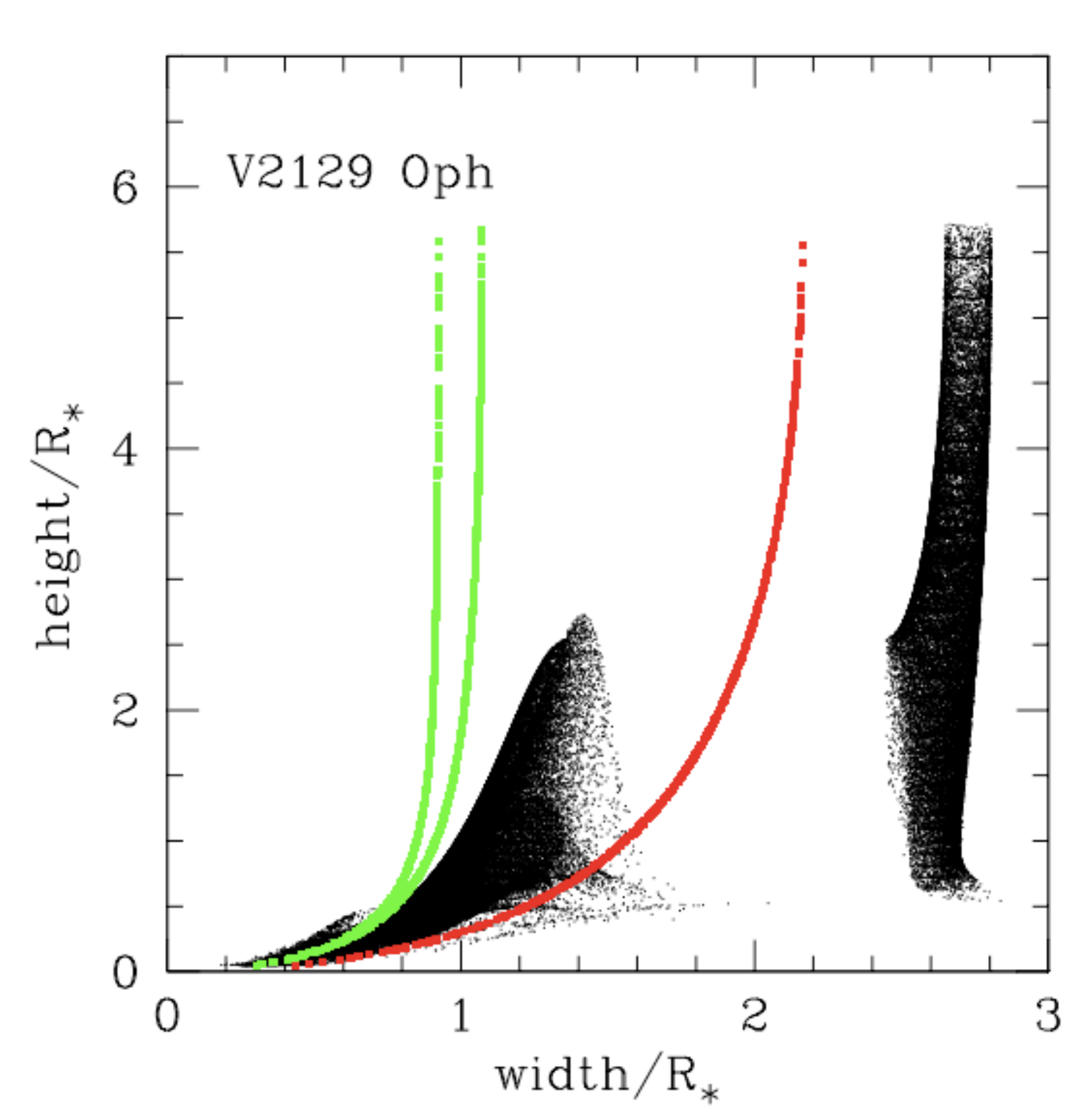}
                        &
                 \label{bptau_height_width}
                 \includegraphics[height=.28\textheight]{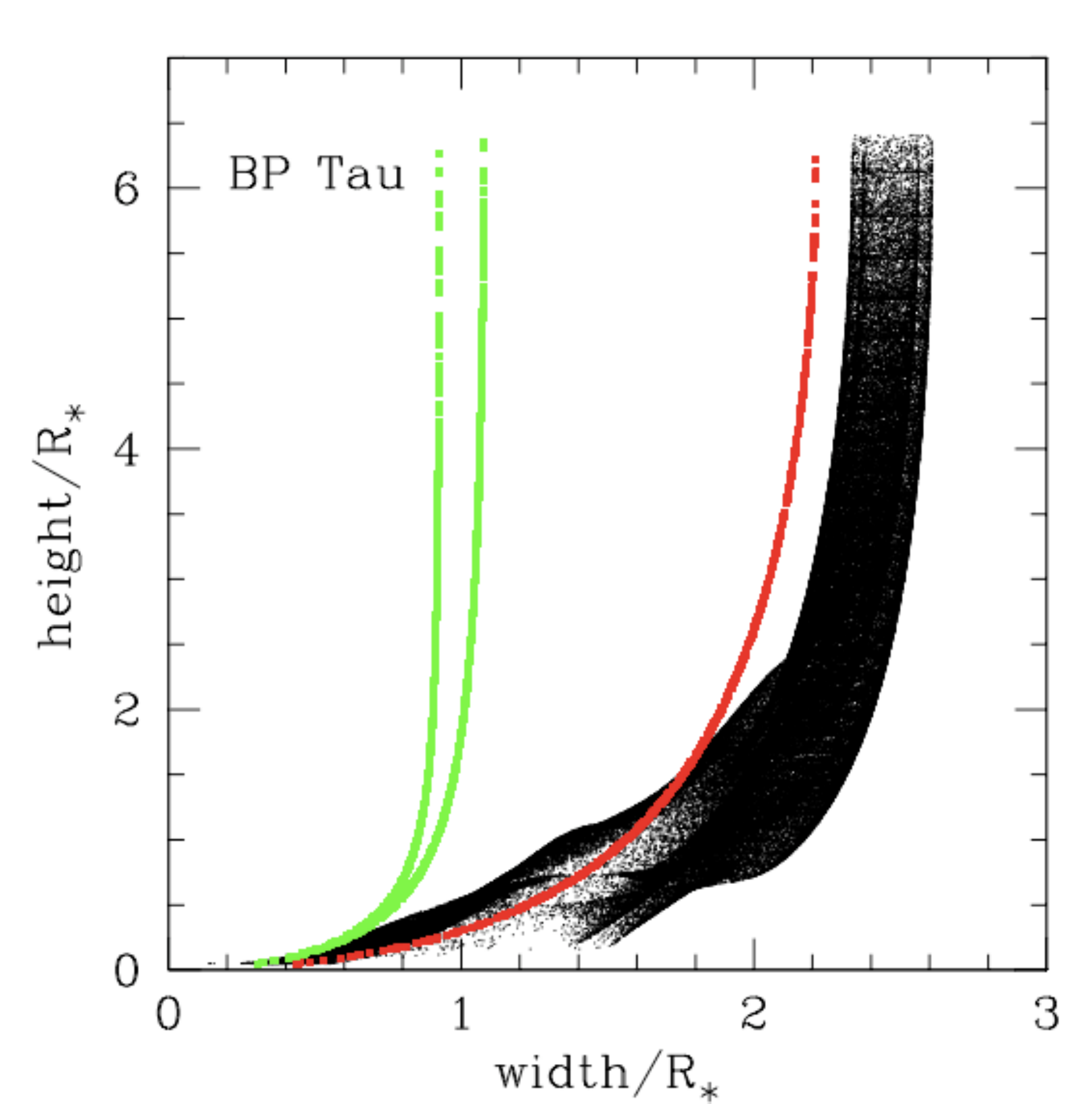}
                          \\
        \end{tabular}
        \caption{Height vs width of the field line loops derived by extrapolation form the surface maps of 
                V2129~Oph (left) and BP~Tau (right).  The black points represent the extrapolated fields and the red/green points a dipole/octupole
                fields.  For both stars, the larger scale field lines are wider than a dipole.}
        \label{height_width}
\end{figure*}

The field extrapolation technique applied specifically to cTTs is discussed in detail by \cite{gre06},
while Fig. \ref{extrap} shows the field topology obtained by extrapolation from the magnetic 
maps of V2129~Oph and BP~Tau shown in Fig. \ref{maps}.   For both stars the surface field 
is complex and non-dipolar, while the larger scale 
field appears more well-ordered (Fig. \ref{extrap}).  As magnetospheric accretion models often assume 
that cTTs have simple dipolar fields (with the exception of recent work, e.g. \cite{lon08}), it is of 
interest to compare the field structures obtained by field extrapolation to a dipole with a source surface.  
Furthermore, as both stars have strong octupole
field components, we also compare the extrapolated fields with octupoles.  To compare the field structures
we calculate the heights and widths of the field line loops obtained by field extrapolation (Fig. \ref{height_width}).  
We define the height, $h$, of a field line loop as the maximum height of the loop above the stellar surface.
The width of a field line, $w$, is defined as the distance along the segment of the great circle connecting the 
field line footpoints on the stellar surface \cite{gre08}.

For BP~Tau the field structure follows a similar trend to a dipole field, which reflects its very strong dipole
field component.  However, the field structure of V2129~Oph shows significant departures from a dipole, 
and is characterized by numerous compact loops.
The difference in field complexity is likely related to the different internal structure of both stars,
which affects the magnetic field generation process, and ultimately the topology of the large scale field.  V2129~Oph is
believed to have a radiative core \cite{don07} and has a particularly complex field structure, whereas BP~Tau is completely convective 
and is found to have a simpler, although still non-dipolar, field topology \cite{don08}.        

For both V2129~Oph and BP~Tau we find that larger scale field lines are wider than dipole magnetic 
loops (Fig. \ref{height_width}).  The larger scale loops, which are the ones that will interact
with the disk, are distorted close to the 
star by the strong surface field regions.  Therefore, even though the large scale magnetic field
is well-ordered and dipole-like, the structure of the accreting field is influenced by the
complex field regions close to the star \cite{gre08}.  We note that recently \cite{fis08} have come (independently) to the same
conclusion for some stars in their sample, and comment that in order to explain the deep absorption 
component detected in a certain line of He, ``flows near the star with less curvature than a dipole trajectory 
seem to be required''.


\subsection{Disk truncation}
\begin{figure*}
        \centering
        \begin{tabular}{cc}
                 \label{v2129oph_trunc_plot}      
                 \includegraphics[height=.28\textheight]{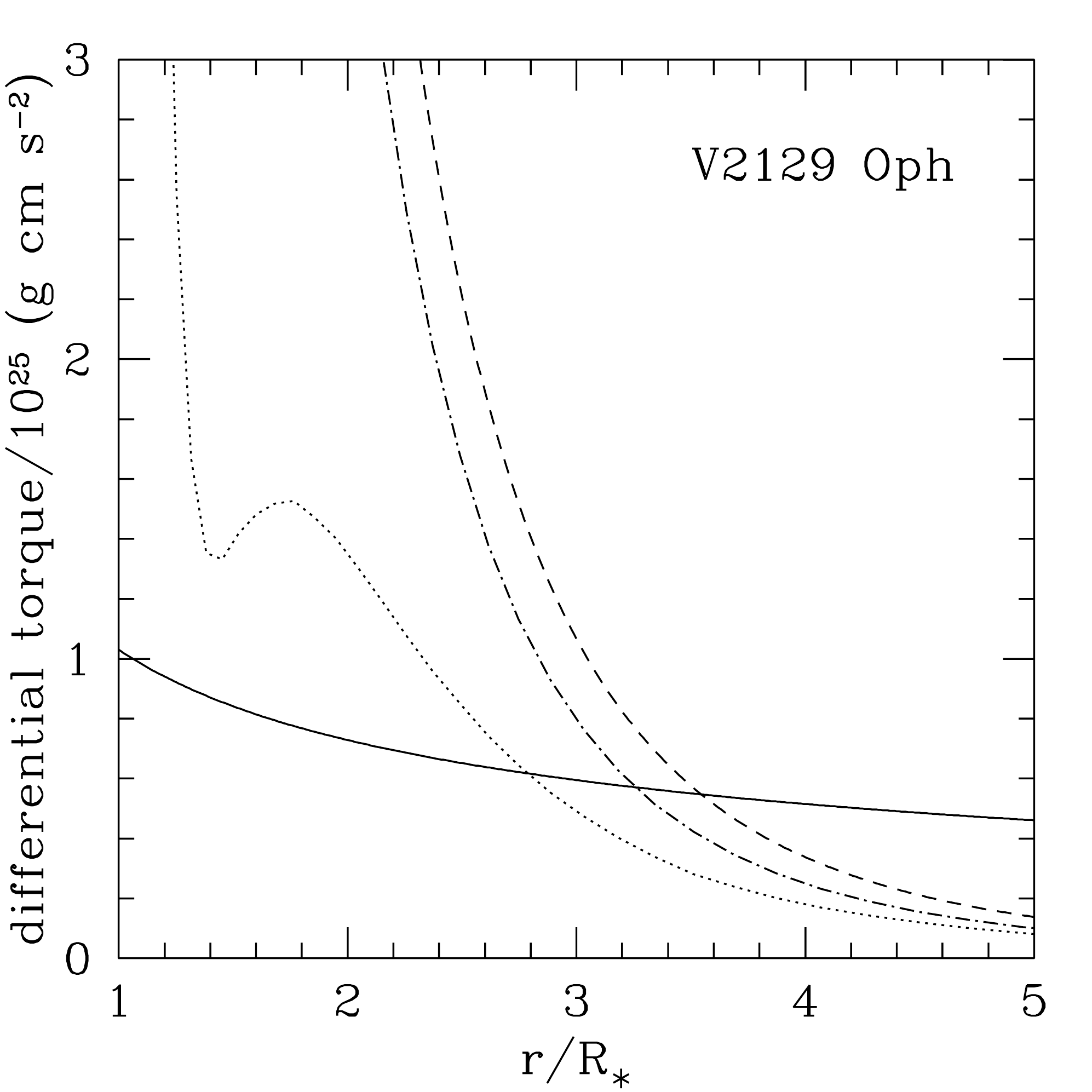}
                        &
                 \label{bptau_trunc_plot}
                 \includegraphics[height=.28\textheight]{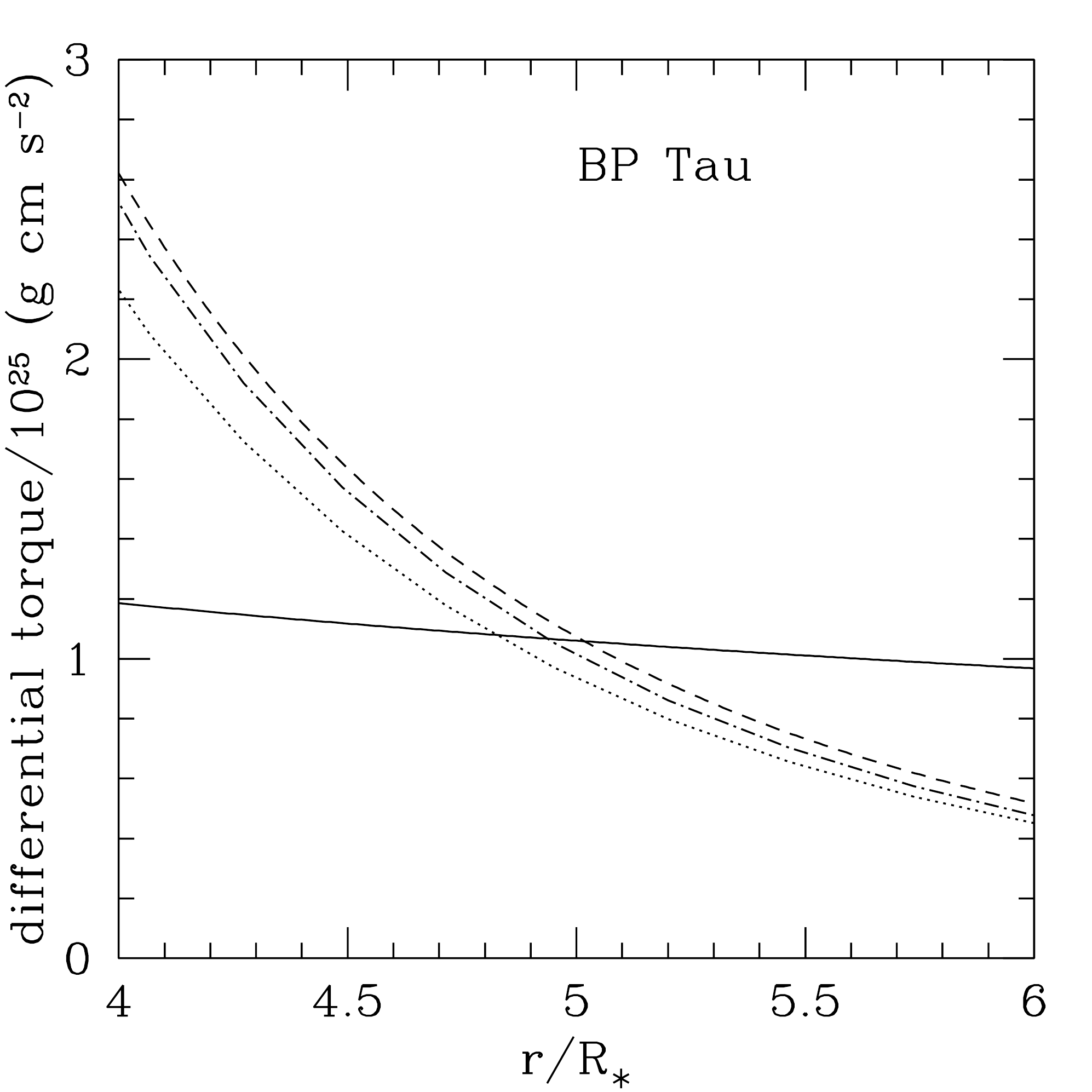}
                          \\
        \end{tabular}
        \caption{The variation along the equatorial plane of the differential magnetic/viscous torque [i.e. the LHS and RHS of equation (\ref{trunc})] 
               for V2129~Oph (left) and BP~Tau (right). The solid line represents the RHS of (\ref{trunc}), with the LHS for an aligned 
               dipole field (dashed line), a tilted dipole field with a source surface (dash-dot line) and the field extrapolation of V2129~Oph/BP~Tau 
              (dotted line).   In both cases the disk is truncated within the corotation radius of 
               $6.7\,{\rm R}_{\ast}$ (V2129~Oph) and $7.4\,{\rm R}_{\ast}$ (BP~Tau).}
        \label{trunc_plot}
\end{figure*}

In this section we compare how the disk truncation radius $R_t$ differs between considering a dipole magnetic
field and the fields derived by field extrapolation.  Various methods of calculating $R_t$ have been 
developed \cite{bes08}, and in this work we assume that the inner disk is truncated where 
the torque due to viscous processes in the disk is comparable to the magnetic torque due to the stellar 
magnetosphere \cite{cla95}.  We assume that at the inner disk the perturbed toroidal component is equal to the  
poloidal component ($B_{\phi} \sim B_z$, see \cite{gre08}).  This assumption 
should be valid as long as the disk is truncated sufficiently within the corotation radius and assuming the disk is
strongly coupled to the stellar magnetic field \cite{mat05b}.  Equating the differential magnetic and 
viscous torques, and assuming that the poloidal field threading the disk is dominated by the vertical
component ($B_z \gg B_r$), gives
\begin{equation}
r^2B_z^2 = \frac{1}{2}\dot{M}\left (\frac{GM_{\ast}}{r} \right )^{1/2}.
\label{trunc}
\end{equation}
The solid lines in Fig. \ref{trunc_plot} represent the variation in the 
RHS equation (\ref{trunc}) along the stars equatorial plane.  We use the same stellar parameters as in \cite{gre08}, namely for
BP~Tau a mass, radius, rotation period and accretion rate of 
$0.7{\rm M}_{\odot}, 1.95{\rm R}_{\odot}, 7.6{\rm d}$ and $2.88\times 10^{-8}{\rm M}_{\odot}{\rm yr}^{-1}$
respectively, and for V2129~Oph, 
$1.35{\rm M}_{\odot}, 2.4{\rm R}_{\odot}, 6.53{\rm d}$ and $1\times 10^{-8}{\rm M}_{\odot}{\rm yr}^{-1}$.  
The various dashed/dotted lines in Fig. \ref{trunc_plot} represent the 
LHS of equation (\ref{trunc}) assuming different forms for the stellar magnetic field.    
The value of $r$ where the lines cross, is then
the disk truncation radius $R_t$.  In the figures we have considered both an analytic dipole field, and a dipole
field with a source surface surface, in order to allow a better comparison with the extrapolated fields
of V2129~Oph and BP~Tau.  We assume that the dipoles have the same polar strength as the measured dipole
components of each star (0.35/1.2~kG for V2129~Oph/BP~Tau).  

For both V2129~Oph and BP~Tau we find that the disk is truncated within the corotation radius, justifying our assumption above (Fig. \ref{trunc_plot}).  
For the extrapolated field of BP~Tau, $R_t$ is only slightly within what is expected for a dipole magnetic field.  This reflects the strong dipole
component of BP~Tau, which is the dominant field component at the inner edge of the disk.  In contrast, $R_t$ for the extrapolated field of V2129~Oph is 
almost a stellar radii within that calculated using a dipole magnetic field.  This is due to intrinsic complexity of the magnetic field of V2129~Oph.  The field 
strength in the disk midplane for this star is less, as the drop-off in field strength with height above the star
is greater than with BP~Tau, and therefore the disk is able to extend closer to the stellar surface than would be expected with  a dipole magnetic field
\cite{gre08}.  Our conclusion is that when considering magnetic fields with a realistic degree of complexity the disk will be truncated
at, or within, the radius determined from dipole magnetic field models.  


\section{Conclusions and future work}
We tentatively find that the complexity of the magnetic field topology of accreting T Tauri stars depends on the internal structure
of the star.  V2192~Oph, which has already developed a radiative core despite its young age, has a more complex and dominantly 
octupolar field, compared to the much simpler field of the completely convective BP~Tau, which has a strong dipole field component.
However, this conclusion is based on only two stars and more data across a range of stellar masses will be required to confirm if this 
is a general feature of all cTTs, although a similar result has been found for low-mass main-sequence stars 
(see Morin et al, these proceedings).  For V2129~Oph there is a more rapid drop-off in field strength with height above the stellar 
surface compared with BP~Tau, which means that the inner disk is truncated much closer to the star compared with a dipole magnetic field
model.  For BP~Tau, with its stronger dipole component, $R_t$ is roughly the same as expected with a dipole field.
Thus the interaction between the stellar field and the disk may be different for stars with different internal structures.

The surface fields of both stars are particularly complex, especially for V2129~Oph, but for both stars the larger scale, likely
accreting field, is more well-ordered.  However, the structure of the larger scale field is distorted close to the stellar surface by the strong
surface field regions.  Thus many of the larger scale field lines are wider than dipole magnetic loops, a result which other authors have
recently found \cite{fis08}. 

In order to quantify the above effects and discover if they are general features of all accreting T Tauri stars, more spectropolarimetric 
data of stars across a range of masses, radii, rotation periods and accretion rates are required.  Such new data will be obtained over the coming
years as part of the MaPP (Magnetic Protostars $\&$ Planets) project, a large program with the ESPaDOnS spectropolarimeter at the 
Canada-France-Hawai'i telescope (PI: J.-F.~Donati), which will provide 690 hours of observing between late 2008 and 2012.   
Such new data will provide an unparalleled opportunity to investigate magnetism on solar-like pre-main sequence stars, and give
new insight into the history of our Sun and Solar System at a time when the planets were begining to form.


\bibliographystyle{aipprocl}

\end{document}